# Direct laser ablation of 2D material films for fabricating multi-functional flexible and transparent devices

*Yigit Sozen[1,*], Yu Kyoung Ryu[2,3], Javier Martinez[2,4], Andres Castellanos-Gomez[1,*]*

[1] *2D Foundry research group. Instituto de Ciencia de Materiales de Madrid (ICMM-CSIC), Madrid, E-28049, Spain.*

[2] *Instituto de Sistemas Optoelectrónicos y Microtecnología, Universidad Politécnica de Madrid, Av. Complutense 30, 28040 Madrid, Spain*

[3] *Departamento de Física Aplicada e Ingeniería de Materiales, E.T.S.I Industriales, Universidad Politécnica de Madrid, C/ José Gutiérrez Abascal 2, 28006 Madrid, Spain*

[4] *Departamento de Ciencia de Materiales-CIME, E.T.S.I Caminos, Canales y Puertos, Universidad Politécnica de Madrid, C/ Profesor Aranguren s/n, 28040 Madrid, Spain*

*corresponding authors: yigit.sozen@csic.es, andres.castellanos@csic.es

ABSTRACT

We present a scalable method for direct patterning of graphite and transition metal dichalcogenide (TMD) films on polycarbonate (PC) and other transparent substrates using fiber laser ablation. This process facilitates the fabrication of various functional devices, including strain gauges, supercapacitors, and photodetector arrays, without the need for photolithography or solvents, thereby simplifying device production and enhancing environmental sustainability. Utilizing roll-to-roll mechanical exfoliation, homogeneous nanosheet films are created and then patterned with a laser engraving system. Electrical and optical characterization confirms that the laser-processed films maintain their crystallinity, with no observable damage to the underlying substrate. We demonstrate the scalability of this approach by constructing a $WSe_2$/graphite photodetector array on PC, which exhibits high sensitivity, low noise, and uniform photocurrent response across its active channels. As a proof-of-concept, this array is used as an image sensor to capture light patterns, showcasing its potential for flexible and semi-transparent imaging applications. These findings open up new avenues for incorporating all-van der Waals devices into wearable electronics, optoelectronics, and imaging technologies.

**INTRODUCTION**

The rapid advancement of flexible and transparent electronics demands efficient, cost-effective fabrication techniques that simplify production processes while maintaining high device performance[1–6]. Conventional patterning techniques, such as photolithography and etching, often require complex steps, expensive equipment, and hazardous chemicals, making them unsuitable for rapid prototyping, scalability, and integration into flexible substrates. In response to these challenges, there is growing interest in developing simpler, more sustainable approaches for direct patterning of materials[7–10].

Laser-based techniques have emerged as promising alternatives for the direct patterning of two-dimensional (2D) materials, offering precision, high processing speed, and environmental benefits[11,12]. Laser-induced graphene (LIG), in particular, has gained significant attention due to its straightforward fabrication process, which involves direct laser writing on carbon-rich substrates to produce porous graphene structures[13,14]. This method enables the creation of intricate patterns without the need for masks or lithography, facilitating rapid prototyping and scalability[15–17]. Recent advancements have expanded LIG production to eco-friendly substrates, such as paper and wood, promoting sustainable electronics manufacturing[18,19]. Additionally, LIG has been integrated into various applications, including energy storage devices and sensors, demonstrating its versatility and functional potential[20,13].

Beyond graphene, laser-assisted synthesis and patterning have been applied to other 2D materials, such as transition metal dichalcogenides (TMDCs)[7,21–24]. Techniques like femtosecond laser ablation allow for precise patterning with single-layer accuracy, enabling the fabrication of complex device architectures[25]. Laser-induced phase transitions and doping have also been employed to modify the properties of 2D materials, enhancing their suitability for electronic and optoelectronic applications[26]. However, despite these advancements, challenges remain in achieving uniformity, scalability, and effective integration of laser-patterned 2D materials into functional devices[27–29].

In this study, we present a laser ablation-based technique for the direct-write patterning of van der Waals (vdW) materials, which addresses many of the challenges associated with traditional and state-of-the-art patterning methods. By employing a pulsed infrared laser for ablation, rapid, top-down patterning of both semiconducting (e.g., $MoS_2$, $WSe_2$) and conducting (graphite) materials on polycarbonate (PC) and other low-absorption substrates was achieved. Unlike previous approaches that focus on either semiconducting or conducting materials, this method demonstrates the ability to pattern multiple vdW materials using the same laser setup, thereby enabling the fabrication of all-van der Waals devices. This integration capability is a key advancement over existing techniques, as it allows for seamless fabrication of multifunctional devices without requiring additional lithography or masking processes.

The laser ablation process, with its precision and high processing speed, enables the creation of complex device structures without the need for photolithography. Importantly, this method is based on an inexpensive laser engraving system and is solvent-free, minimizing the environmental concerns related to chemical usage and waste, which are inherent in many traditional patterning methods. Laser patterning approach is complemented by a high-throughput, roll-to-roll-like mechanical exfoliation process that facilitates the preparation of homogeneous nanosheet films, allowing for scalable production[30].

The versatility of the technique is further demonstrated by its capacity to combine different vdW materials into multifunctional device architectures. Devices such as strain gauges, supercapacitors, and photodetectors were seamlessly integrated on flexible substrates. Notably, the direct-write approach offers a unique capability for rapid prototyping, which is crucial for the iterative development of novel devices in both research and industrial settings. The ability to fabricate conducting and semiconducting regions using a single process distinguishes the present work from laser-induced graphene techniques that primarily focus on graphitizing carbon-based materials[13,31]. By directly patterning pre-deposited graphite films, the challenges posed by graphite's high thermal conductivity are overcome, enabling precise and efficient patterning through pulsed laser ablation.

Furthermore, we explore the scalability of the laser ablation method by fabricating a $WSe_2$/graphite photodetector array and demonstrating its potential as an image sensor for light pattern detection. The photodetector array, fabricated on PC, shows high sensitivity, low noise, and uniform response, making it a promising candidate for integration into future wearable and adaptable imaging systems. This versatility, combined with the simplicity and environmental benefits of the approach, positions it as a significant advancement in the field of 2D materials, providing a scalable and cost-effective route for the fabrication of next-generation flexible and transparent electronics.

**RESULTS AND DISCUSSIONS**

**Fabrication and laser ablation of the films**

**Fabrication of nanosheet films:**

We used a high-throughput, roll-to-roll-like mechanical exfoliation method to produce nanosheet films of van der Waals materials[30]. By continuously exfoliating thin layers from bulk crystals using rolling cylinders equipped with Nitto SPV 224 tape, we ensured homogeneous film formation on an acceptor tape. The exfoliation duration and the number of re-exfoliations depended on the material type (e.g., graphite, $MoS_2$, $WSe_2$). Different materials required varying rolling times to achieve homogeneous films. For example, transition metal dichalcogenides (e.g., $MoS_2$, $WSe_2$) typically needed 30-40 seconds for a single exfoliation, resulting in a uniform distribution. In contrast, graphite often formed thick, large crystallites during initial exfoliation, reducing transfer yield. To address this, the acceptor tape containing crystallites was replaced with clean Nitto tape for re-exfoliation. This re-exfoliation process minimized crystallite formation, resulting in a more homogeneous distribution of flake sizes and thicknesses, thereby recovering high transfer yields and ensuring continuous film integrity. Optical microscopy images illustrated in Figures S1 show the tapes after each exfoliation step and the subsequent transfer of graphite films to PC substrates. Image in Figure S2 represents the final condition of each tape after transfer of graphite films on PC. As it is seen, after the fourth exfoliation, high transfer yield was achieved, which will further enable consistent and continuous flake network formation for subsequent transfers.

**Thermal release transfer process:**

The transfer process involved adhering the tape to an acceptor substrate and annealing at 100°C for 5 minutes. During this heating process, the Nitto SPV 224 tape loses its adhesion, allowing the flakes to be naturally transferred onto the acceptor substrate surface. Interestingly, recent XPS studies have shown that this process does not leave adhesive residues on the surface of the transferred flakes. XPS, being highly sensitive to surface contamination, confirmed the cleanliness of the transferred films[32].

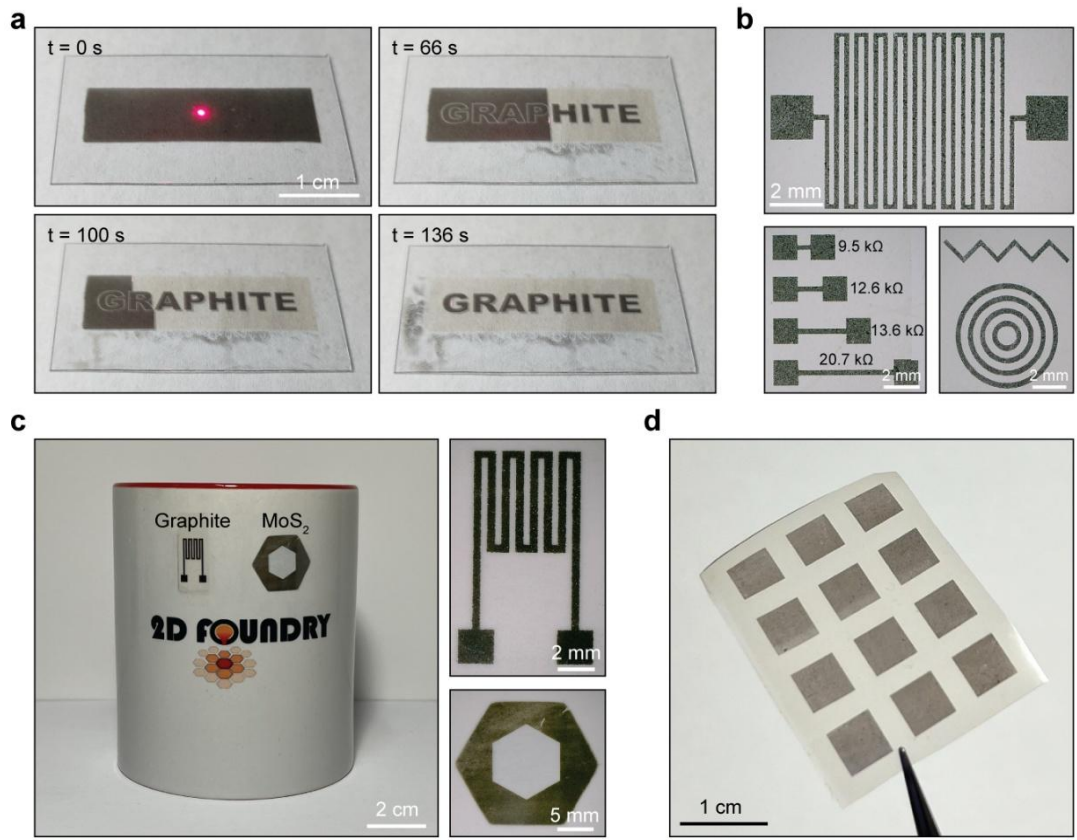

**Figure 1. Laser ablation patterning of graphite, MoS$_2$, and WSe$_2$ films on various substrates.** (a) Sequential images of the laser ablation process on a roll-to-roll exfoliated graphite film deposited on a PC substrate, with times indicating the elapsed duration from the start of the ablation. (b) Laser-patterned graphite structures on a glass substrate, including a meander-line inductor (top), resistors with different geometries and resistances (bottom left), and sawtooth and spiral shapes (bottom right). (c) The left image shows the overall view of graphite and MoS$_2$ patterns on a ceramic substrate, while the right images are the magnified views of a strain gauge (top) and a hexagonal structure (bottom), illustrating compatibility with curved surfaces. (d) Patterned WSe$_2$ film on Nitto tape after roll-to-roll exfoliation.

**Laser ablation for direct write patterning and versatility across different substrates:**

We employed a pulsed infrared laser (1064 nm) engraving system from Atomstack, model M4, to pattern 2D films on flexible PC substrates, glass substrates, or directly on the Nitto SPV 224 tape. Figure 1a illustrates the patterning of a large-area graphite film (fabricated by two transfer steps) on a flexible PC substrate using the laser engraving system to remove the graphite from the desired locations by direct laser ablation. The final image demonstrates the ability of the laser ablation method to achieve clear and precise patterns of 2D films onto a flexible substrate. Remarkably, the entire film area is patterned in just 136 seconds, which is significantly faster than other widely used traditional techniques such as etching and photolithography. This method proved substantially faster and more efficient for scalable manufacturing. This rapid processing capability demonstrates the method's suitability for scalable, cost-effective manufacturing, rapid prototyping, and integration into semiconductor fabrication. Moreover, the laser ablation process is environmentally friendly, requiring no solvents for post-processing; only a nitrogen gun is needed to remove dust from the ablation region. Figure 1b shows the laser-patterned meander line inductor, resistors with varying channel lengths, and geometric shapes on a glass substrate, demonstrating the versatility of the method in creating diverse structures. We also patterned graphite and $MoS_2$ films directly transferred on a ceramic mug (Figure 1c) to create a strain gauge and a hexagon-shaped pattern, respectively, illustrating the adaptability of the approach to curved surfaces. Moreover, this method enables the patterning of as-prepared 2D films on Nitto tape. Figure 1d shows square-patterned $WSe_2$ films on Nitto tape, which were successfully transferred onto substrates such as $SiO_2/Si$ (see Figure S3). Direct laser patterning on $SiO_2/Si$ is challenging due to the high optical absorption of the Si layer, which leads to overheating and substrate damage during laser exposure. Using pre-patterned films that can be then transferred onto $Si/SiO_2$ allows us to overcome these limitations effectively. Given the critical role of laser type and parameters in substrate interactions, future studies could investigate alternative laser systems optimized to minimize thermal effects in Si-based substrates[33,34].

**Structural and electronic characterization of patterned films**

**Raman characterization of the laser ablation processed materials:**

To assess the crystallinity of the film following the laser ablation process, spatial Raman mapping was conducted in a region adjacent to the laser-ablated edge. A reflection mode micrograph of the selected mapping area, covering 0.01 mm$^2$, is shown Figure S4a. As observed, there are no visual structural changes between the inner and edge flakes, indicating the absence of mechanical damage beyond the ablation zone. Prior to mapping, full Raman spectrum were collected from three distinct locations, (i) the graphite film, (ii) the edge, and (iii) ablated substrate (see Figure S4b), to highlight the variations in Raman features across the ablated region. The Raman spectrum obtained from the film region displays G and 2D peaks of graphite, located at 1579 cm$^{-1}$ and 2722 cm$^{-1}$, respectively[35–37]. The strong intensities of these peaks, along with the absence of D peak (defect-induced and localized around 1350 cm$^{-1}$), indicate high crystallinity at the selected point. In contrast, D peak becomes measurable in the edge region, which can be attributed to an increase in defect density or structural disorders induced by laser ablation. Additionally, substrate-related features were observed, including a background signal and a Raman peak at 3075 cm$^{-1}$, corresponding to C-H bond stretching in benzene rings[38] arising from the PC substrate underneath. To assess potential damage to the PC substrate during the patterning process, Raman spectra of both ablated and pristine PC substrates were acquired. Both spectra show identical Raman peaks without any noticeable peak shift (see Figure S4b), indicating no significant alteration in polymer crystallinity[39,40]. The peak positions for all spectra are summarized in Table S1.

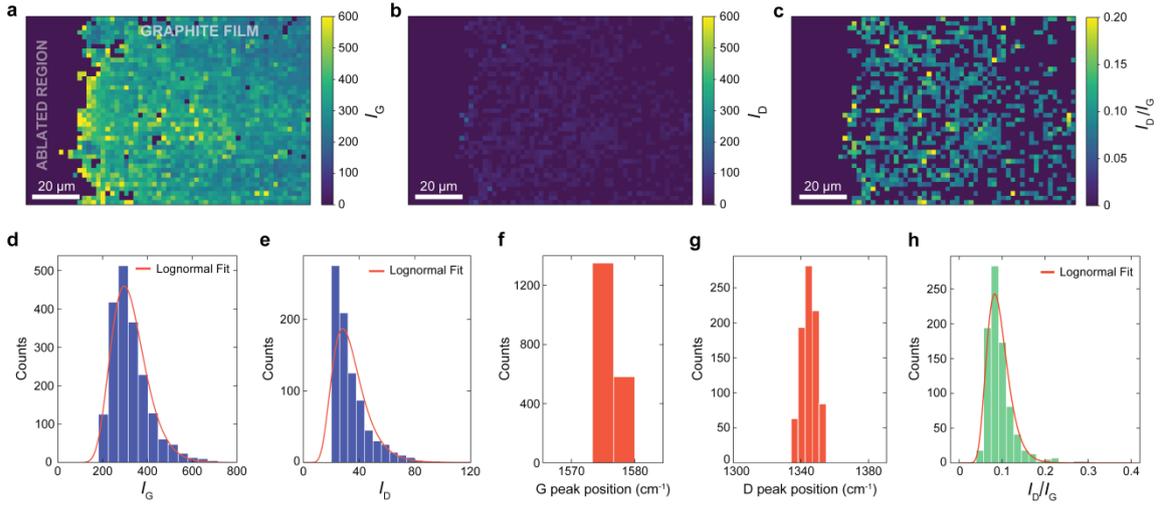

**Figure 2. Raman mapping of the graphite film near the laser-ablated edge**. (a–c) Raman maps showing the spatial distribution of G peak intensity ($I_G$), D peak intensity ($I_D$), and the calculated $I_D/I_G$ intensity ratio across the mapped region, respectively. The Raman map reveals a strong G peak signal throughout the film region, while the D peak signal is significantly lower, indicating a high degree of crystallinity. (d-h) Histograms showing the distribution of $I_G$, $I_D$, G peak position, D peak position, and $I_D/I_G$ ratio, respectively.

Subsequently, a Raman map was acquired with a spatial resolution of 2 µm to provide detailed structural information across a large film area. Figure 2a and 2b show the spatial distribution of the G peak intensity ($I_G$) and D peak intensity ($I_D$), respectively. The strong and continuous presence of G peak over the entire film region confirms the maintenance of graphitic structure after laser ablation. In contrast, the D peak is significantly less prominent compared to the G peak and is absent in certain locations. Although a slight increase in the D peak intensity is observed at the edge, the difference relative to the inner region of the film remains minimal, indicating that the laser process does not substantially impact the film's crystallinity. The $I_D/I_G$ map in Figure 2c further supports these findings, owing to low ratio values (especially in the regions further from the ablated edge) indicating high structural order and low defect density across the film[41–43].

Histograms presented in Figure 2d-h provide quantitative insights into Raman spectral parameters including peak intensities, positions and intensity ratios. The histograms in Figure 2d and 2e reveal right-skewed distributions for $I_G$ and $I_D$, for which the modes were calculated to be 297 and 28, respectively, based on lognormal function fitting. The G and D peak positions (Figure 2f and 2g) show narrow distribution, centered around ~1574 cm$^{-1}$ and ~1345 cm$^{-1}$, respectively. The position of the Raman peaks was reported to be dependent on strain and charge doping[44]. Therefore, a narrow distribution means that the whole film is uniform in terms of strain and doping, with no effect from the edge ablation. Notably, the $I_D/I_G$ distribution (Figure 2h) also follows a narrow lognormal profile, with a mode at 0.08, indicating low defect density and confirming the maintenance of high crystallinity in the graphite film following laser ablation.

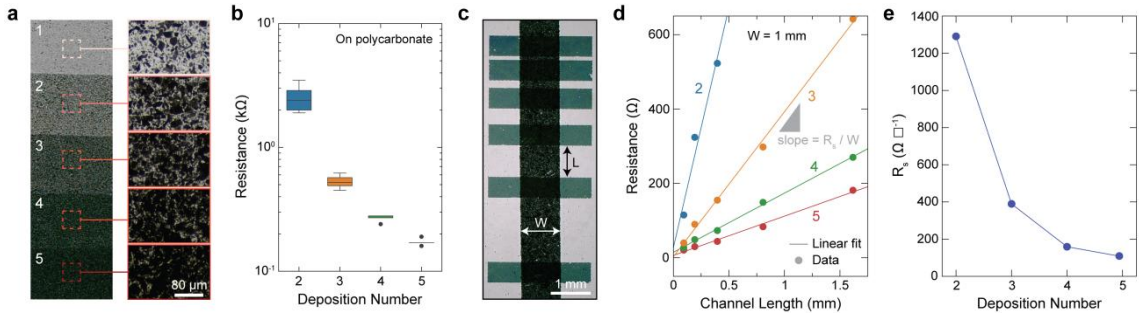

**Figure 3. Electrical characterization of graphite films with increasing transfer layers on PC substrate.** (a) Left: Optical microscope images in transmission mode showing the development of a continuous graphite film on PC as the number of film transfers increases from 1 to 5. Right: High-magnification images of selected regions (indicated by dashed squares) reveal increased film density and uniformity with additional layers. (b) Box plots of resistance measurements across graphite films with varying number of transfers, demonstrating reduced resistance with more transferred steps. (c) Configuration of a sample prepared for transfer length method (TLM) analysis, consisting of a 5-transfer graphite film on PC with a patterned width of 1 mm and multiple gold (Au) contacts evaporated on top at varying inter-electrode distances. (d) Transfer length method analysis showing resistance as a function of channel length for films with 2 to 5 transfers. Linear fits (slope = $R_s$/W) are applied to calculate sheet resistance and contact resistance of each film. (e) Sheet resistance ($R_s$) as a function of the number of deposition steps, indicating a progressive decrease in $R_s$ with additional transfers, reflecting improved conductivity upon nanosheets stacking.

**Film conductivity as a function of depositions steps:**

To achieve a well-conductive network, we improved the electrical conductivity of the interconnected graphite flakes by increasing the number of film depositions. Each additional layer created more percolative pathways, enhancing the electrical transport properties. Figure 3a, left, shows optical microscopy images of graphite films on PC, formed by increasing the number of deposited layers. As observed, homogeneous and well-covered films were achieved by stacking several layers on top of each other.

Higher magnification images reveal that subsequent transfers lead to densely packed films, effectively minimizing empty spots (Figure 3a, right).

Figure 3b presents box plots that illustrate the variability of resistance across films with increasing deposition numbers. To create these box plots, we collect resistance data from various regions of each film using two-probe resistance measurements, maintaining a constant distance of ~2 mm. The central box represents the interquartile range, with the black vertical line inside indicating the median value. The whiskers extending from the box show the upper and lower extreme values, while the data with diamond shape indicate outliers. The percolation threshold was reached after just two depositions, indicating the formation of long conductive channels through overlapping graphite flakes. As the number of conductive channels increased with additional layers, a rapid decrease in film resistance was observed. Moreover, the standard deviation in resistance decreased, reflecting the improved uniformity of the flake network. Homogeneous conductivity throughout the entire graphite film (~2.5 cm$^2$) was achieved after four or five depositions.

**Transfer length method measurements:**

Figure 3c shows an example of a sample prepared for transfer length method (TLM) measurements, comprising five deposited graphite layers. The graphite film was patterned into a bar shape with a width (*W*) of 1 mm, followed by the evaporation of an array of gold (Au) contacts with varying inter-electrode spacings. Figure 3d displays the resistance values as a function of channel length for samples with different numbers of deposited layers. A decreasing trend in resistance was observed as the channel length between the Au contact pairs decreased. By fitting the data to a linear function, we extracted both the contact resistance ($R_c$) and sheet resistance ($R_s$) of the films. The y-intercept of the fitting line corresponds to $2R_c$, while the slope represents $R_s/W$.

Contact resistances were calculated as 13.8, 1.5, 7.4, and 3.2 Ω for films with two to five depositions, respectively. The calculated sheet resistances are shown in Figure 3e, decreasing from 1291 Ω/□ to 106 Ω/□ as the number of film depositions increased. A sharp drop in sheet resistance from 2 to 3 transfers indicates that flakes start to form a continuous, interconnected network. Then it remains nearly constant from 4 to 5 transfer, implying that the network is now well-established. The resistivity ($\rho$) of the film was derived from the thickness ($t$) and the sheet resistance ($R_s$) using the formula $\rho = R_s t$. For a graphite film with five depositions, we obtained an average thickness of 158 nm using atomic force microscopy (Figure S5a and S5b). Moreover, Figure S5c presents height profiles of three distinct locations to assess the surface morphology and thickness variations across the edge region. Accordingly, the resistivity $\rho$ was estimated to be $1.7 \cdot 10^{-5}$ Ω m, consistent with previous studies that report in-plane resistivity values for bulk ($\approx 10^{-5} - 10^{-6}$ Ω m)[45] or nanosheet graphite films fabricated with electrochemical exfoliation method ($\approx 1.7 \cdot 10^{-5}$ Ω m)[46]. Resistivity can be converted to conductivity ($\sigma$) by using the equation of $\rho = 1/\sigma$. The calculated $\sigma$ is $5.8 \cdot 10^4$ S/m, which is higher than most reported values for graphite films fabricated through liquid phase exfoliation methods[47–54], even though there are few studies reporting values higher than this value[55–57].

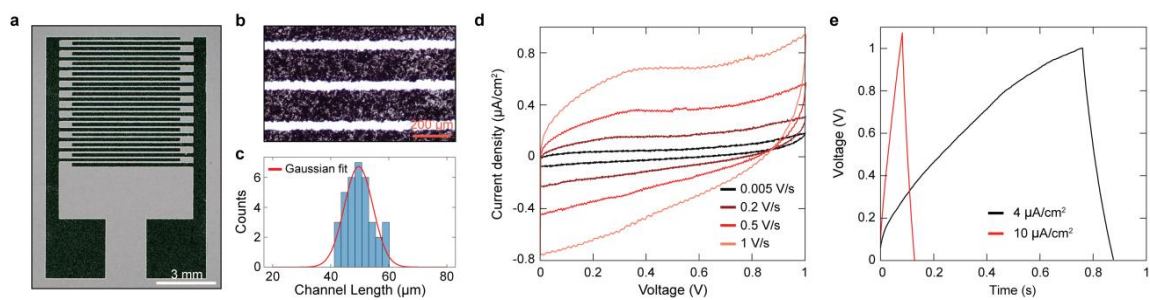

**Figure 4. Characterization of interdigitated graphite electrodes on PC for micro-supercapacitor applications.** (a) Optical image of the interdigitated graphite electrode pattern on a PC substrate, showing the overall electrode design with closely spaced "fingers" for enhanced surface area. (b) High-magnification optical image in transmission mode displaying the uniformity and separation of the graphite fingers. (c) Histogram of the channel lengths (gap between electrode fingers) in the interdigitated graphite electrode structure. Gaussian distribution curve (red line) gives an approximate mean of 50 μm for the channel length. (d) Cyclic voltammetry (CV) curves of the micro-supercapacitor at different scan rates, demonstrating the device's electrochemical performance across a range of voltage sweep speeds. (e) Galvanostatic charge-discharge (GCD) curves measured at two current densities (4 μA/cm$^2$ and 10 μA/cm$^2$), illustrating the charge storage behavior and stability of the microsupercapacitor under different operational conditions.

**Graphite micro-supercapacitor proof-of-concept**

To illustrate the potential of the laser ablation patterning technique, we fabricated several multifunctional devices by patterning a graphite film (formed by multiple deposition steps). Figure 4a shows an optical microscopy image of one of the fabricated interdigitated electrode devices on a PC surface, while Figure 4b presents a high-magnification image detailing the electrodes and gaps, to be used as a micro-supercapacitor. The fabrication process of the interdigitated electrode is presented in Video S1 (Supporting Information). Figure 4c shows the statistical distribution of the channel length (distance between electrodes), with a narrow distribution centered around 50 μm.

Recent advancements have highlighted the potential of graphite-based materials as low-cost and efficient current collectors for energy storage devices. For instance, graphite tape, fabricated by pressing a Kapton tape onto a graphite sheet, has been demonstrated as a low-cost current collector with excellent electrical conductivity and mechanical flexibility in sandwich supercapacitors utilizing MnO$_2$ as the active electrode material[58]. Similarly, graphite ink and two commercially available graphite foils have been evaluated as low-cost current collectors in sandwich supercapacitors employing activated carbon electrodes[59].

We explored the capability of interdigitated graphite electrode to store energy as in-plane micro-supercapacitor. The fabricated electrodes were coated with a poly(vinyl alcohol) (PVA)-1M $H_2SO_4$ gel electrolyte, and silver paint was applied to the contact pads to ensure reliable electrical connectivity. The electrochemical performance of the resulting devices was characterized using a potentiostat/galvanostat (Autolab PGSTAT204), with all results normalized to the active device area of approximately 0.5 $cm^2$.

The cyclic voltammetry (CV) curves, taken over a potential range of 0–1 V at scan rates ranging from 0.005 to 1 V/s (Figure 4d), exhibit a near-rectangular shape. This indicates typical behavior of an electrical double-layer capacitor, with an increase in the enclosed area of the CV curves (corresponding to stored energy) as the scan rate increases. Similarly, galvanostatic charge-discharge (GCD) measurements conducted at two different current densities (Figure 4e) display triangular shapes, consistent with the behavior expected from electrical double-layer capacitors.

Using the GCD curve recorded at a current density of 4 $\mu A/cm^2$, an areal capacitance of 0.48 $\mu F/cm^2$ was calculated based on the following equation[60]:

$$C_A = \frac{I_{discharge}}{S \times \frac{dV}{dt}}$$

where $I_{discharge}$ is the discharge current, $S$ is the active area of the device (previously noted as 0.5 $cm^2$), and $dV/dt$ represents the slope of the discharge curve.

The calculated areal capacitance of 0.48 $\mu F/cm^2$ is relatively small, as expected, due to the very low porosity of the electrodes, as observed in the scanning electron microscope (SEM) image in Figure S6. Despite this limitation, we have demonstrated that interdigitated graphite electrodes fabricated with roll-to-roll mechanical exfoliation plus laser ablation patterning exhibit capacitive behavior, which can contribute to the energy storage mechanism as an active material when employed as a low-cost, high conductive current collector in supercapacitors.

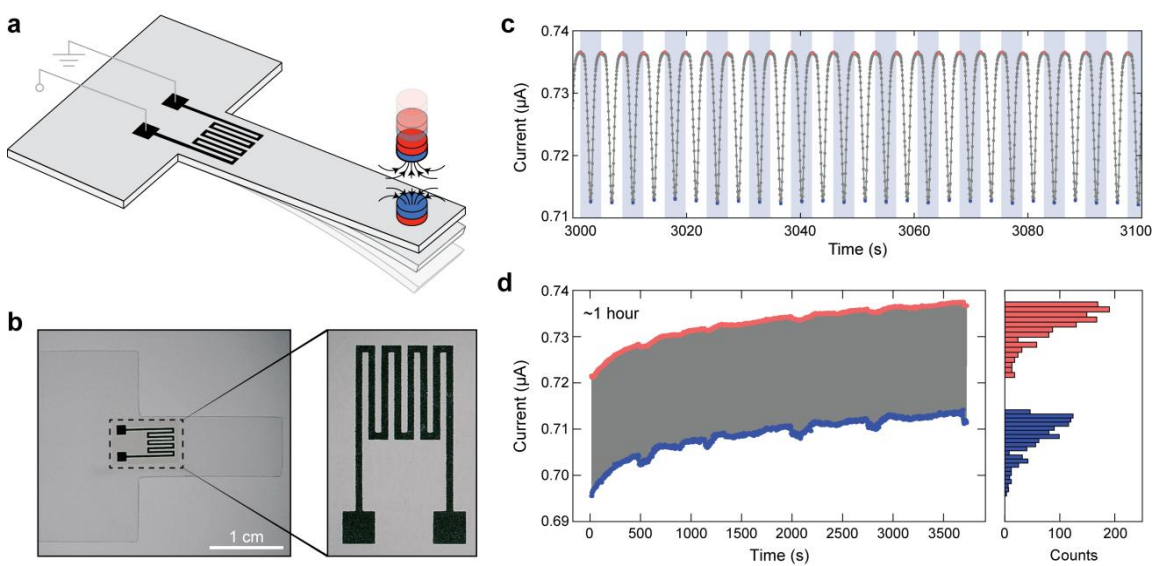

**Figure 5. Strain sensing performance of a graphite-based strain gauge on PC cantilever.** (a) Schematic illustrating the experimental setup for applying external force to a PC cantilever with an integrated graphite strain gauge structure, using magnets to induce cyclic strain. (b) Optical images of the cantilever, showing the strain gauge fabricated by laser-patterning of a graphite film. (c) Time-resolved current measurement demonstrating the strain gauge's reversible response to cyclic loading (blue-shaded regions indicate periods of applied strain, and white regions indicate release), highlighting the gauge's sensitivity to mechanical deformation. (d) Long-duration stability test (~1 hour) of the strain gauge under repetitive strain cycles, with the plot on the right showing the distribution of current values during strained (blue) and relaxed (red) states, indicating consistent performance over time despite the slow drift which could be attributed to changes in temperature/humidity along the measurement.

**Mechanical durability of a graphite strain gauge**

Next, we conducted a strain test on a graphite strain gauge patterned onto a PC cantilever to evaluate the mechanical durability of the device under cyclic deformation. The cantilever, equipped with a magnet at one end, was clamped on a stage, and cyclic bending was induced using a linear motor stage that moved another magnet back and forth to deflect the free end of the cantilever (Figure 5a). Figure 5b presents optical images of the cantilever with the integrated strain gauge.

The strain gauge's response under continuous bending cycling was monitored by performing time-resolved current measurements while biasing the device with 1 V. Figure 5c illustrates the variation in current as the cantilever was cyclically strained and released. The observed decrease in current during straining can be attributed to an increase in overall resistance, which arises from the separation between interconnected graphite flakes leading to a reduction of the percolative conduction pathways[46,61].

To assess the long-term stability of the strain gauge, the current was continuously monitored for approximately one hour during repeated strain cycles (Figure 5d). The results show that while the current changes consistently during each strain cycle, a gradual drift in the baseline current was observed over time, amounting to approximately 1.4 %. This drift could potentially be attributed to changes in environmental conditions such as temperature or humidity during the measurement period.

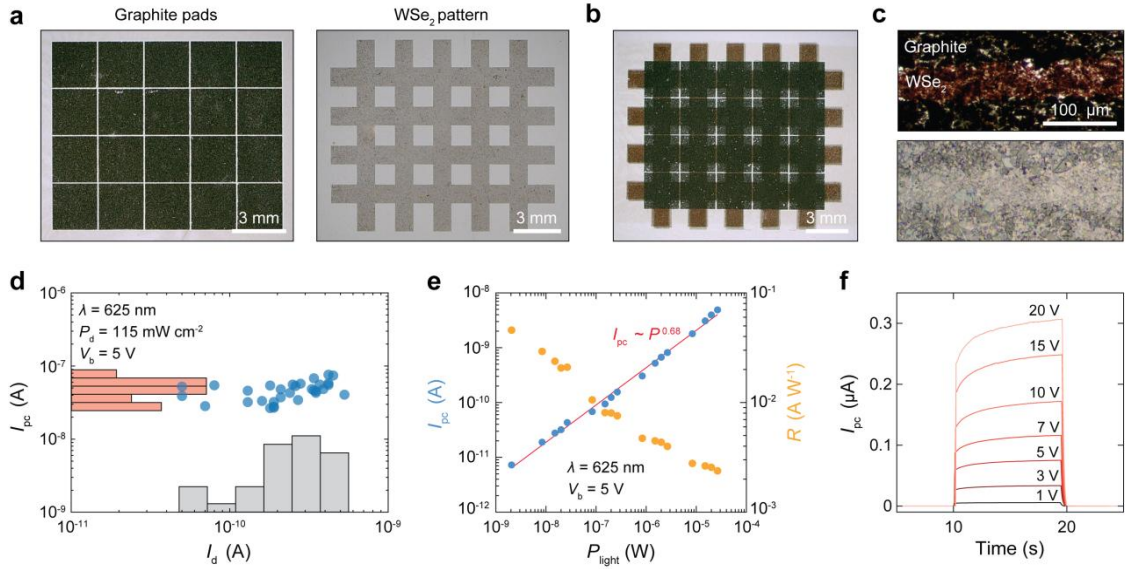

**Figure 6. Fabrication and characterization of a WSe$_2$/graphite-based photodetector array on PC substrate.** (a) Laser patterned graphite pads on PC substrate (left) and WSe$_2$ film on Nitto tape (right). (b) Optical image of the completed WSe$_2$/graphite photodetector array, comprising 31 optically active WSe$_2$ channels formed by transferring 5 films of WSe$_2$ onto the graphite pads. (c) Reflection and transmission mode micrographs of the active channel, showing the bridging of the graphite electrodes by WSe$_2$ flake networks, forming the photodetector channel. (d) Histogram illustrating the distribution of photocurrent ($I_{pc}$) and dark current ($I_d$) values across individual pixels of the array, highlighting the uniformity and reproducibility of the photodetector response. (e) Plot of photocurrent ($I_{pc}$) and photoresponsivity (R) as functions of illumination power ($P_{light}$), with a power-law fit (red line) demonstrating the device's sensitivity to light intensity. (f) Photoswitching characteristics under different bias voltages, with the light source ($\lambda = 625$ nm, $P = 0.73$ mW) turned ON at $t = 10$ s for a duration of 10 s.

**Fabrication and imaging performance of WSe$_2$/graphite photodetector array**

To demonstrate the scalability and potential of integration of the laser ablation method, we fabricated an array of photodetectors by combining graphite and WSe$_2$ films. First, graphite film was deposited onto a PC substrate by performing five sequential film transfers, resulting in a uniform graphite structure. This film was then patterned using the laser to define a 4×5 array of square graphite electrodes, with each electrode separated by a 50 µm gap to act as a channel region (see Figure 6a, left). Separately, we prepared a mesh-like structure of WSe$_2$ film on Nitto tape, forming a grid pattern with WSe$_2$ bars, ensuring that its dimensions matched the electrode layout (see Figure 6a, right). This patterned WSe$_2$ film was then transferred onto the pre-patterned graphite electrode array using a thermal release process. During the transfer, the WSe$_2$ grid was carefully aligned so that it lay precisely on top of the graphite pads. This transfer process was repeated five times to ensure the formation of continuous semiconducting nanosheet network between neighboring graphite electrodes across the array. As each channel between the electrodes serves as a unit photodetector, this design allows for the formation of a photodetector array comprising a total of 31 pixels, causing the formation of two sets of arrays which can be readout as $4 \times 4$ and $5 \times 3$. (see Figure 6b). Transmission and reflection micrographs for one of the channels (Figure 6c) demonstrate a uniformly covered channel with well-distributed WSe$_2$ flakes.

Figure 6d presents the measured photocurrent ($I_{pc}$) values as a function of the dark current ($I_d$) for each pixel, along with histograms showing photocurrent and dark current distributions. We determined the photocurrent values by measuring the photoswitching characteristics of each pixel through time-resolved current measurements while the illumination source is switched ON and OFF in periods of 10 seconds. The measurements were conducted under a bias voltage ($V_b$) of 5 V, and a 625 nm LED light source with a power density ($P_d$) of 115.2 mW cm$^{-2}$. The dark current values were obtained by calculating the average of the data measured in the dark condition. Most of the measured photocurrent values fall within the same order of magnitude, indicating that the photodetector array exhibits a uniform optical response. The photocurrent values vary from 27 to 76 nA, clustering around 50 nA. Each pixel shows low dark current values ranging from $10^{-10}$ to $10^{-9}$ A, resulting with high signal-to-noise ratios ($I_{pc}/I_d$) ranging between $10^2$ and $10^3$, which reflects high sensitivity with minimal noise.

We extended our measurements by investigating the photocurrent characteristics of one of the pixels under varying light power. Figure 6e presents the measured photocurrent values obtained from time-resolved current measurements along with the corresponding photoresponsivities as a function of increasing light power. Photoresponsivity was extracted using the equation:

$$R = \frac{I_{pc}}{P} \times \frac{A_{spot}}{A_{effective}}$$

where $I_{pc}$ is the photocurrent, $P$ is the excitation power, $A_{spot}$ is the area of the focused light spot, and $A_{effective}$ is the effective active channel area between the electrodes. $A_{spot}$ is 0.64 mm$^2$, and $A_{effective}$ is 0.05 mm$^2$, which was calculated by multiplying the total channel area with a factor $c$ (0 completely empty channel and 1 completely covered by flakes), in order to obtain the area just covered with flakes. For the considered photodetector unit, $c$ was obtained as 0.9 which indicates a 90 % of coverage for the channel. The photocurrent shows a sub-linear dependence on the incident light power, which can be expressed using the power law $I_{pc} \sim P^{\alpha}$, where α is the power exponent that characterizes the response. According to the power law fit to the photocurrent versus light power data, was calculated to be 0.68, suggesting the presence of a photogating effect in the system. The photodetector remained responsive at light power as low as ~ $10^{-9}$ W, indicating its capability to detect very weak optical signals. The lowest light power excitation yielded a responsivity of 0.05 A/W, higher than that of photodetectors based on 2D films fabricated via liquid phase exfoliation methods, which typically range from 0.001 to 28 mA/W[62–69].

Figure 6f shows the fast photoswitching behavior of the photodetector unit under different bias voltages ($V_b$). As the bias voltage increased, the carrier separation in the semiconducting channel improved, leading to a gradual increase in the photocurrent from 6 to 303 nA when the bias voltage was swept from 1 V to 20 V. Stable photocurrent characteristics up to 20 V demonstrate the device's reliability over a wide operational range. We used 10 %-90 % method in order to obtain the response times for the rise and decay sides of the photocurrent signals. Figure S7 provides the evaluation of rise ($\tau_{rise}$) and decay times ($\tau_{decay}$) with increasing $V_b$. While $\tau_{decay}$ showed no dependence on $V_b$, $\tau_{rise}$ increased up to 10 V, however the trend was interrupted beyond this threshold due to an underestimation of $\tau_{rise}$, attributed to an insufficient ON cycle duration, preventing the photocurrent from reaching saturation.

To further demonstrate the potential application of the WSe$_2$/graphite photodetector array as an image sensor, a patterned light source was projected onto the array and measured the photocurrent response for each pixel. For this, we considered 16 pixels forming the 4x4 array for image acquisition. Using a collimated LED light source ($\lambda$ = 625 nm, $P_d$ = 0.58 mW cm$^{-2}$), we created an "H" pattern by placing a laser-cut aluminum foil mask in the light path, positioned over a bi-convex lens to project the shape onto the photodetector array (Figure 7a). This setup enabled controlled illumination of specific regions on the array, allowing us to evaluate the array's capability to resolve spatial light patterns. Figure 7b shows an optical image of the "H" pattern projected over the array, where the illuminated and dark regions are clearly distinguishable. Correspondingly, the photocurrent response at each pixel was recorded to generate a heat map of the array (Figure 7c). The heat map reveals photocurrent values that align with the projected pattern, with higher values in the illuminated regions and lower values in the shaded areas. This response illustrates the uniform sensitivity and spatial resolution of the photodetector array. Moreover, we demonstrated the consistency of the photodetector array in image sensing by successfully detecting a distinct light pattern (see Figure S8). The experiment demonstrates the scalability and potential of the laser-patterned $WSe_2$/graphite photodetector array for basic imaging applications, where its high signal-to-noise ratio and sensitivity to weak light signals can facilitate the detection of spatially-resolved light distributions. The photodetector array's fabrication on a PC substrate not only showcases the versatility of the laser ablation process but also opens avenues for developing flexible and semi-transparent image sensors, paving the way for applications in wearable electronics, augmented reality, and lightweight, adaptable imaging systems.

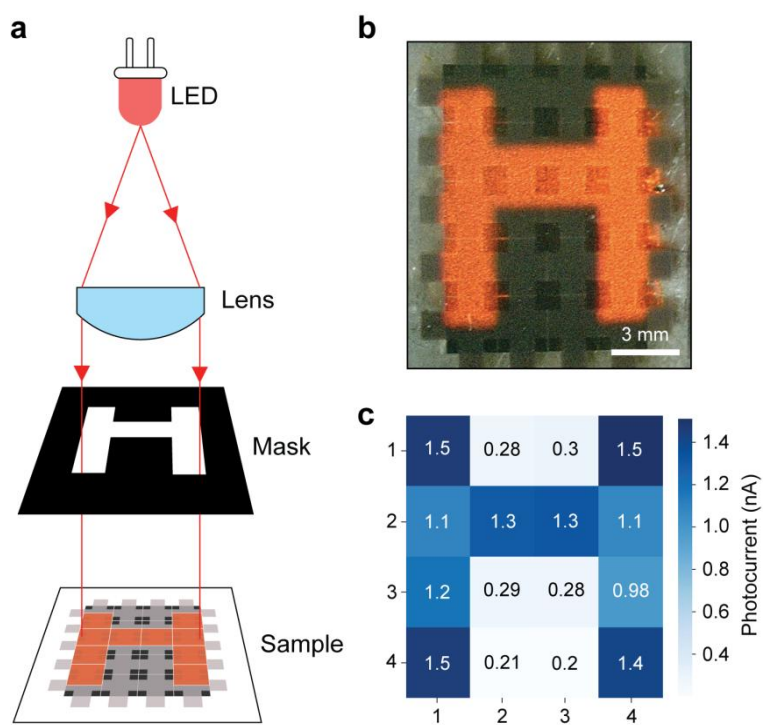

**Figure 7. Application of the $WSe_2$/graphite photodetector array as an image sensor.** (a) Schematic of the projection setup used for image detection, consisting of an LED light source, a bi-convex lens, and a laser-cut

aluminum foil mask forming an "H" pattern. This setup projects the pattern onto the surface of the photodetector array. (b) Optical image of the photodetector array with the projected "H" pattern. (c) Heat map of the photocurrent response measured across the array, with values (in nanoamperes) representing the detected light intensity at each pixel. This response corresponds to the projected pattern, illustrating the potential of the photodetector array for basic imaging applications by resolving light intensity distributions across the sensor array.

**CONCLUSION**

In this work, we developed and characterized a laser ablation method for directly patterning graphite, $MoS_2$ and $WSe_2$ films on PC and other low-absorption substrates, establishing an efficient approach for creating all-van der Waals devices without additional lithography steps. This method enables the scalable fabrication of high-performance devices, including strain gauges, supercapacitors, and photodetectors, with precise control over pattern geometry and minimal environmental impact. Notably, the fabricated $WSe_2$/graphite photodetector array on PC demonstrated uniform optical sensitivity, low dark current, and high signal-to-noise ratios, making it suitable for flexible and semi-transparent image sensing applications. The successful demonstration of an image sensor based on this photodetector array underscores the versatility of the approach for future applications in flexible electronics, wearable devices, and lightweight imaging systems. The laser ablation technique, in combination with high-throughput exfoliation, paves the way for integrating 2D materials into advanced electronic and optoelectronic devices, contributing to the development of sustainable, adaptable, and high-performance technologies.

MATERIALS AND METHODS

Crystals and Substrates:

Natural graphite flakes were purchased from ProGraphite GmbH. $MoS_2$ exfoliation was performed by using bulk natural molybdenite mineral supplied from Molly Hill Mine, Quebec, Canada. $WSe_2$ single-crystal flakes were purchased from HQ Graphene. Polycarbonate films with 250 μm thickness were supplied from Modulor (www.modulor.de) to be used as flexible substrates. Microscope glass slides were used as a glass substrate.

Laser Ablation of Two-Dimensional Films:

We used laser engraving system from Atomstack (model M4) to pattern 2D films fabricated with high-throughput roll-to-roll exfoliation. The pattern designs were prepared using SeaCAD sofware, which also offers precise control over engraving parameters such as laser power, speed, and frequency. The patterning process of films was carried out using 15% of the laser's total power (total power is 20 W) with a constant speed of 50 mm/s, ensuring controlled ablation without damaging the substrate. Other parameters were kept as default as represented in Figure S9.

Electronic and Optical Measurements:

Electronic and optoelectronic characterizations were conducted at room temperature using a homebuilt probe station. Time-resolved current measurements were carried out using a Keithley 2450 source-meter unit. A fiber-coupled LED source with a wavelength of 625 nm (Thorlabs M625F2) was employed to perform bias- and power-dependent photocurrent measurements. The LED source was operated in modulation mode to adjust the output light power by varying the current through an external voltage supplier (TENMA 72-2715). In image sensing application, we used bi-convex lens (Thorlabs LB1471) to focus light pattern. Light is patterned with aluminum foil masks, which were prepared using the laser engraving system set to 100% power and a constant speed of 1 mm/s. The light power was measured using an optical power meter (Thorlabs, PM100D) coupled to a photodiode power sensor (Thorlabs, S120C).

Raman Characterization:

Raman mapping was performed under ambient conditions using a confocal Raman microscope (MonoVista CRS+, Spectroscopy & Imaging GmbH). The optical excitation was provided by a 532 nm line of a continuous wave (CW) solid-state laser with an incident light power of 0.32 mW through a 50x magnification microscope objective (NA = 0.75). The 300 lines/mm grating were used. For the Raman spectra shown in Figure S4b, an incident laser power of 0.72 mW was used, focused through a 100x microscope objective (NA = 0.9).

Scanning electron microscope (SEM) measurement:

Secondary electron images were taken using a scanning electron microscope (FEI Inspect F50) under an accelerating voltage of 1 kV to characterize the surface of the graphite electrode.

Atomic Force Microscopy:

A commercial Atomic Force Microscopy (AFM) system, from Nanotec, operating in ambient conditions was employed to perform the height characterization. Measurements have been acquired in dynamic mode with silicon tips (PPP-FMR from Nanosensors). Image analysis was performed with Gwyddion free software[70].

DATA AND CODE AVAILABILITY

Any additional information required to reanalyze the data reported in this paper is available from the lead contact upon reasonable request.

ASSOCIATED CONTENT

**Supporting Information:**

Supplementary Information includes:

Figure S1 Optical microscope images of re-exfoliated graphite films on Nitto and after transfer on PC substrate

Figure S2 Final condition of graphite films after transfer

Figure S3 Transfer of square-patterned $WSe_2$ films on $SiO_2$/Si substrate

Figure S4 Optical microscopy image and site-specific Raman characterization of the laser-ablated graphite film

Figure S5 AFM image and height profiles of a graphite film on PC

Figure S6 SEM image of the graphite flake network in an interdigitated electrode

Figure S7 Bias-dependent rise and decay times in a $WSe_2$/graphite photodetector

Figure S8 Distinct light pattern sensing with $WSe_2$/graphite photodetector array

Figure S9 Image showing engraving parameters used in laser ablation of the films

Table S1 Measured Raman modes for pristine and ablated PC, and their comparison with literature

AUTHOR INFORMATION

**Corresponding Authors**

Yigit Sozen yigit.sozen@csic.es;

Andres Castellanos-Gomez andres.castellanos@csic.es

**Author Contributions**

**Yigit Sozen**: Data curation (lead); Formal analysis (lead); Investigation (lead); Methodology (lead); Writing-original draft (lead); Writing-review & editing (equal).

**Yu Kyoung Ryu**: Data curation (supporting); Formal analysis (supporting); Investigation (supporting); Methodology (supporting).

**Javier Martinez**: Data curation (supporting); Formal analysis (supporting); Investigation (supporting); Methodology (supporting).

**Andres Castellanos-Gomez**: Conceptualization (lead); Funding acquisition (lead); Methodology (supporting); Project administration (lead); Resources (lead); Supervision (lead); Writing-original draft (supporting); Writing-review & editing (equal).

ACKNOWLEDGEMENTS

We thank Dr. Carmen Munuera (ICMM-CSIC) for her support with the AFM measurements and useful discussions along the work. We also would like to acknowledge ICTS Micronanofabs. This work was funded by the Ministry of Science

and Innovation (Spain) through the projects TED2021-132267B-I00, PID2020-115566RB-I00, PRE2021-098348, PID2023-151946OB-I00, and PDC2023-145920-I00. The authors also acknowledge funding from the European Research Council (ERC) through the ERC-PoC 2024, StEnSo project (grant agreement no. 101185235), the FLAG-ERA program (JTC 2019) under the project To2Dox (PCI2019-111893-2), the Comunidad de Madrid through the CAIRO-CM project (Y2020/NMT6661), and the CSIC program for the Spanish Recovery, Transformation and Resilience Plan funded by the Recovery and Resilience Facility of the European Union, established by the Regulation (EU) 2020/2094.